\documentstyle[osa,manuscript]{revtex}

\begin{document}

\title{Laser cooling and trapping of Yb from a thermal 
source}
\author{Umakant D. Rapol, Anusha Krishna, Ajay Wasan, and 
Vasant Natarajan\thanks{Electronic address: 
vasant@physics.iisc.ernet.in}}
\address{Department of Physics, Indian Institute of 
Science, 
Bangalore 560 012, INDIA}

\maketitle
\begin{abstract}
We have successfully loaded a magneto-optic trap for Yb 
atoms from a thermal source without the use of a Zeeman 
slower. The source is placed close to the trapping region 
so that it provides a large flux of atoms that can be 
cooled and captured. The atoms are cooled on the ${^1S_0} 
\leftrightarrow {^1P_1}$ transition at 398.8 nm. We have 
loaded all seven stable isotopes of Yb into the trap. For 
the most abundant isotope ($^{174}$Yb), we load more than 
$10^7$ atoms into the trap within 1 s. For the rarest 
isotope ($^{168}$Yb) with a natural abundance of only 
0.13\%, we still load about $4 \times 10^5$ atoms into 
the trap. We find that the trap population is maximized 
near a detuning of $-1.5\Gamma$ and field gradient of 75 
G/cm.
\end{abstract}

\section{Introduction}

Laser-cooled atoms open exciting new possibilities for 
several experiments including those in precision 
spectroscopy. Yb is a particularly good choice for such 
studies because of several reasons. Yb has been proposed 
for experiments in quantum optics \cite{HCT87,BZW95} due 
to the existence of two nearly-closed low-lying excited 
states with widely differing lifetimes. Laser-cooled Yb 
has been proposed for optical frequency standards 
applications \cite{HZB89}. Yb has been shown to be a 
useful candidate for the test of electroweak interactions 
because there is an enhanced parity-violation effect in 
the ${^1S_0} \leftrightarrow {^3D_1}$ transition 
\cite{DEM95}. Laser-cooled Yb extends the possibilities 
in measuring the strength of parity-violating 
interactions with unprecedented accuracy due to the 
availability of dense samples where one can apply large 
electric fields and homogeneous magnetic fields. 
Furthermore, Yb has a wide range of stable isotopes, both 
fermionic and bosonic, allowing one to explore the 
possibility of studying the properties of fermion-boson 
mixtures \cite{MFS99,CZT00}.

The starting point of most laser-cooling experiments is 
the magneto-optic trap (MOT) \cite{RPC87}. The source of 
atoms for the MOT is usually an oven inside the vacuum 
chamber containing a small amount of pure metal which is 
heated to release a beam of atoms. For atoms with high 
melting point such as Na, the hot atoms emanating from 
the oven are first slowed by a counter-propagating laser 
beam (Zeeman slower or chirp slower). Yb has a melting 
point of 819$^{\circ}$C and previous experiments have 
used a Zeeman slower for loading the MOT 
\cite{HTK99,LBS00}. However, we have earlier shown that a 
MOT for Rb atoms can be loaded efficiently from a thermal 
getter source operating at temperatures near 
300$^{\circ}$C without the use of a slower \cite{RWN01}. 
Since Yb has considerable vapor pressure over the solid 
at 300$^{\circ}$C, we have been exploring the possibility 
of loading a Yb MOT directly from a thermal source. 

In this paper, we show that this is indeed possible by 
successfully loading all the seven stable isotopes of Yb 
into the MOT from a thermal source. For the most abundant 
isotope ($^{174}$Yb), we load about $1.7\times10^7$ atoms 
with a time constant of 190 ms. Most significantly, we 
can trap the rarest isotope ($^{168}$Yb) which has a 
natural abundance of only 0.13\%. There are two factors 
that make this possible. The first is that the oven is 
placed close to the trapping region so that it provides a 
large flux of atoms that can be captured by the MOT. The 
second factor is that the excited state lifetime of the 
cooling transition is only 5.68 ns and the photon 
scattering rate is very high. Atoms are slowed over a 
short distance and small laser beams of 10 mm diameter 
are enough to cool and capture atoms from the background 
vapor. 

\section{Experimental details}

Yb has two possible cooling transitions, as seen from the 
energy level diagram in Fig.\ 1. The ${^1S_0} 
\leftrightarrow {^3P_1}$ intercombination line at 555.8 
nm forms a pure two-level system in Yb due to the absence 
of ground-state hyperfine structure. The transition has a 
long lifetime of 874 ns corresponding to a natural 
linewidth of $\Gamma = 2\pi \times 182$ kHz, and 
therefore the photon cycling rate is very slow. However, 
the Doppler cooling limit is 4 ${\mu}$K, a very desirable 
temperature for high-precision measurements. The other 
cooling transition is the ${^1S_0} \leftrightarrow 
{^1P_1}$ transition at 398.9 nm. This has a short 
lifetime of 5.68 ns or a linewidth of $\Gamma = 2\pi 
\times 28$ MHz, and allows photons to be cycled quickly. 
However, it has the disadvantage of giving a high Doppler 
cooling limit of 670 $\mu$K, and is not a completely 
closed transition since the excited state can branch into 
the low-lying $^3P$ levels through the intermediate $^3D$ 
levels. Despite these disadvantages, we have decided to 
use the 398.9 nm transition for our experiments primarily 
because the high photon scattering rate allows atoms to 
be cooled over a short distance.

The laser light for the 398.9 nm transition is produced 
in a two-step process. First, the output of a tunable 
Ti-sapphire laser (Coherent 899-21) is set to half the 
desired frequency near 798 nm. Then its output is fed 
into an external-cavity frequency-doubling unit (Laser 
Analytical Systems, LAS100) which uses a nonlinear 
crystal of lithium triborate to produce the desired 398.8 
nm light. The Ti-sapphire laser is actively stabilized 
using an ovenized, Fabry-Perot reference cavity and its 
instantaneous linewidth is less than 1 MHz. We have also 
measured that its long-term drift is only a few MHz per 
hour, which makes it unnecessary to lock the laser to an 
external reference. For our experiments, we find that it 
is sufficient to set the frequency at the desired point 
and leave it for the duration of the experiment. The 
frequency is set using a home-built wavemeter 
\cite{BRW01} that uses a scanning Michelson 
interferometer and a Rb-stabilized reference laser. The 
wavemeter has an absolute accuracy of a few MHz which is 
more than sufficient for finding the Yb lines. Once the 
trap is working, the trap fluorescence signal is used to 
optimize the laser frequency or detuning.

The experiments are done inside a vacuum chamber 
consisting of a standard six-way cross made of pyrex with 
25-mm O.D. windows. The pyrex cell is pumped by a 20 l/s 
ion pump and maintained at a pressure below $10^{-9}$ 
torr. The source of Yb atoms is a quartz ampoule 
containing a small amount of elemental Yb. The ampoule is 
a 30 mm long tube of 5 mm diameter that is sealed at one 
end and constricted to 1 mm diameter at the other end. 
The constriction seems important both in ensuring a 
uniform flux in the trapping volume and in preventing the 
walls from getting coated with a film of Yb. Using a 
collimated thermal beam, as is the case when using a 
Zeeman slower, causes the opposite wall to get rapidly 
coated with a thick film of Yb. On the other hand, we 
find no such deposit even after 50 hours of operation of 
the source. The source is placed about 50 mm from the 
trap center. It is resistively heated to a temperature of 
300--400$^{\circ}$C to release Yb.

The MOT is formed from a standard configuration 
consisting of three pairs of mutually orthogonal laser 
beams and a quadrupole magnetic field superposed at the 
intersection of the beams. The total laser power 
available near the MOT is 90 mW. This is split into three 
circularly-polarized beams. Each beam has a power of 30 
mW and diameter of 10 mm. The saturation intensity $I_0$ 
for the transition is 58 mW/cm$^2$, therefore the 
intensity in the beams is about $0.65 I_0$. The beams are 
retroreflected after passing through the cell, and the 
intensity in the return beams is slightly smaller due to 
losses along the optical path. The quadrupole magnetic 
field is produced using a pair of coils (270 turns each) 
placed 40 mm apart. With 3 A of current through the 
coils, the field gradient at the center is 90 G/cm. There 
is negligible heat generation in the coils at this 
current. The fluorescence from the trapped atoms is 
imaged on to a calibrated photo-multiplier tube (PMT). 
The number of trapped atoms is estimated from the total 
power incident on the PMT after accounting for losses 
along the optical path \cite{RWN01}. There is an error of 
about 20\% in estimating the number of atoms in this 
manner, but relative values between isotopes and other 
such trends are unaffected by this error.

\section{Results and discussion}

In the first set of experiments, we scanned the laser 
slowly across a frequency range of 3 GHz. The 
fluorescence signal detected by the PMT is shown in Fig.\ 
2. The signal shows peaks as each isotope comes into 
resonance and gets trapped in the MOT. The peak heights 
correspond roughly to the natural abundance of each 
isotope except for the odd isotopes which show anomalous 
behavior. This is because the odd isotopes have 
half-integer nuclear spin and consequently have hyperfine 
structure in the $^1P_1$ excited state. In these cases, 
optical pumping into the wrong hyperfine level causes 
additional loss from the trap. The signal from the rarest 
isotope $^{168}$Yb is also seen in the figure. As shown 
in the inset of Fig.\ 2, the $^{168}$Yb peak becomes 
clearer when the gain of the PMT is increased by a factor 
of 100.

In order to better characterize the source and find 
optimal values for the laser detuning, we have studied 
the trap population as a function of laser frequency. The 
laser was scanned slowly over a frequency range of 100 
MHz below the resonance. The results for $^{174}$Yb are 
shown in Fig.\ 3. The number of trapped atoms shows a 
peak value of $1.7\times10^7$ at a detuning of 
$-1.5\Gamma$. This is similar to the behavior seen in the 
case of a Rb MOT loaded from a thermal source 
\cite{RWN01}. Note that the optimal value of detuning 
depends on the size of the trapping beams and the 
magnetic-field gradient, since both these parameters are 
important in determining the distance over which atoms 
are slowed and captured. The data in Fig.\ 3 were taken 
with a magnetic field gradient of 75 G/cm. All the other 
isotopes show similar dependence on detuning.

The loading of the MOT from a thermal source follows a 
rate equation \cite{RWN01}
\begin{equation} 
\label{rateeq} 
\frac{dN}{dt} = R - \frac{N}{\tau} ,
\end{equation}
where $N$ is the number of atoms in the trap, $R$ is the 
rate at which atoms are captured from the source, and 
$\tau$ represents the losses from the trap. In the case 
of Yb, there are several factors that contribute to the 
loss rate. The dominant mechanism is collisional losses, 
both due to collisions with background atoms and 
collisions with hot, untrapped Yb atoms. In addition, as 
mentioned before, the ${^1P_1}$ excited state of the 
cooling transition can branch into the low-lying $^3P$ 
levels. Of these, the ${^3P_0}$ and ${^3P_2}$ levels are 
metastable levels; atoms shelved into these levels no 
longer participate in the cooling cycle and are rapidly 
lost from the trap \cite{LBS00}.

We have studied the loss mechanisms in the trap by 
studying the loading characteristics of the MOT for the 
different isotopes. The solution to the rate equation in 
Eq.\ 1 is an exponential growth in the number of trapped 
atoms: $N = N_s [1- \exp(-t/ \tau)]$, where $N_s$ is the 
steady state population and is given by $R \tau$. The 
loading characteristics of the most abundant isotope 
$^{174}$Yb and the least abundant isotope $^{168}$Yb are 
shown in Fig.\ 4. Both curves follow the exponential 
behaviour predicted from Eq.\ 1. Deviation from the 
exponential curve would indicate other loss meachnisms. 
For example, losses due to collisions between trapped 
atoms would appear as an $N^2$ term in Eq.\ 1 
\cite{APE94}. Under our conditions, such intra-trap 
collisional loss does not seem significant. From Fig.\ 4, 
we see that for $^{174}$Yb the time constant is 190 ms 
and the steady state population is $1.3\times10^7$. For 
$^{168}$Yb, the time constant is 440 ms while the steady 
state population is about 40 times smaller. We should 
also mention that the optimal value of magnetic-field 
gradient for trapping $^{174}$Yb and $^{168}$Yb are 
different, with values of 75 G/cm and 60 G/cm, 
respectively.

The characteristics of the source and the loss mechanisms 
become more apparent by studying the values of $\tau$ and 
$N_s$ for the different isotopes under identical 
conditions. The measured values are listed in Table 1. 
The data were taken at a detuning of $-1.2\Gamma$ and 
field gradient of 75 G/cm. The value of $\tau$ for the 
different isotopes lies in that range of 160--450 ms, 
while $N_s$ varies from $3\times10^6$ to $1.3\times10^7$. 
The general trend is that $\tau$ decreases and $N_s$ 
increases with the natural abundance of the isotope, 
except for the odd isotopes which, as mentioned before, 
show anomalous behavior due to optical pumping in the 
excited-state hyperfine levels. The trends become clearer 
if we consider the values for the loading rate and $N_s$ 
for each isotope normalized to the corresponding value 
for $^{174}$Yb, as listed in Table 2. For the even 
isotopes, the loading rate scales as the natural 
abundance, as expected from a non-enriched source. On the 
other hand, the value of $N_s$ scales roughly as the 
square root of the natural abundance, implying that 
$\tau$ is inversely proportional to the square root of 
the abundance. One possible explanation for this is that 
the cross-section for collisions with same isotope atoms 
are larger than the cross-section for collisions with 
other isotopes.

\section{Conclusions}

We have successfully loaded a MOT for Yb 
atoms from a thermal source without the necessity of a 
Zeeman slower. The source is placed close to the trapping 
region so that it provides a large flux of atoms that can 
be captured by the MOT. We have loaded all seven stable 
isotopes of Yb. For the most abundant isotope 
($^{174}$Yb), we load about $2\times10^7$ atoms into the 
trap within 1 s. Even for the rarest isotope ($^{168}$Yb) 
with a natural abundance of only 0.13\%, we load about 
$4\times10^5$ atoms into the trap. We have studied the 
trap population as a function of laser detuning and find 
that the population is maximized near a detuning of 
$-1.5\Gamma$. We have characterized the losses from the 
trap by studying the loading of the MOT as a function of 
time. We find that the number of trapped atoms grows 
exponentially with a time constant that scales as the 
inverse square root of the natural abundance of the 
isotope. Trap losses due to optical pumping into the 
metastable $^3P_0$ and $^3P_2$ levels could be a 
significant limitation on the number of trapped atoms. In 
future, we plan to clear out these levels using diode 
lasers at 649 nm (exciting the ${^3P_0} \rightarrow 
{^3S_1}$ transition) and at 770 nm (exciting the ${^3P_2} 
\rightarrow {^3S_1}$ transition). With these 
improvements, we expect to get increased population and 
lifetime in the trap which is important for many of the 
proposed experiments using laser-cooled Yb.

\acknowledgments

We thank V. Anandan for fabricating the pyrex cell. This 
work was supported by a research grant from the 
Department of Science and Technology, Government of 
India.

\begin{figure}
\caption{Yb energy levels. The figure shows the relevant 
low-lying energy levels in Yb. Even-parity states are on 
the left and odd-parity states on the right. The two 
possible cooling transitions at 398.9 nm and 555.8 nm are 
shown.
}
\label{yblevels}
\end{figure}

\begin{figure}
\caption{MOT fluorescence vs. laser frequency. The 
fluorescence signal shows peaks as each isotope comes 
into resonance with the laser and is trapped in the MOT. 
Each peak is labeled by the corresponding isotope number. 
The inset shows a $100\times$ magnified view of the 
region where $^{168}$Yb is trapped.
}
\label{nvsiso}
\end{figure}

\begin{figure}
\caption{Number of trapped atoms vs. detuning. The number 
of $^{174}$Yb atoms in the MOT is shown as the detuning 
of the laser is varied slowly from $-60$ MHz ($-2\Gamma$) 
to 0. The magnetic field gradient at the trap center was 
75 G/cm. The number of atoms shows a peak at $-42$ MHz or 
$-1.5\Gamma$. Other isotopes show identical behavior.
}
\label{nvsdet}
\end{figure}

\begin{figure}
\caption{Loading of the MOT. The build up of the number 
of atoms in the MOT is shown after the trapping beams are 
turned on at $t=0$. In (a), we have a MOT for $^{174}$Yb 
at a detuning of $-1.2\Gamma$ and field gradient of 75 
G/cm. In (b), we observe signal from a MOT for $^{168}$Yb 
at a detuning of $-1.2\Gamma$ and field gradient of 60 
G/cm. Both curves show exponential growth in the number 
of atoms with time constants of 190 ms and 440 ms, 
respectively. Note that under steady state, the number of 
trapped $^{168}$Yb atoms is about 40 times smaller.
}
\label{load}
\end{figure}

\begin{table}
\caption{
Listed are the loading time constant $\tau$ and steady 
state population $N_s$ in the MOT for the various 
isotopes. The data for each isotope were taken at a 
detuning of $-1.2 \Gamma$ and field gradient of 75 G/cm. 
The errors in $\tau$ are statistical $1\sigma$ 
deviations. For $N_s$, there is a calibration error of 
about 20\% due to errors in converting the detected 
fluorescence signal to the number of atoms. However, the 
error is the same for all isotopes and does not affect 
relative values.
}
\begin{tabular}{cccl}
Isotope &Nat. abund. & $N_s$ & $\tau$ (ms) \\
\tableline
176 &12.7\% & $8.0\times10^6$ & 300(30) \\
174 & 31.8\% &$1.3\times10^7$ & 190(10) \\
172 & 21.9\% &$1.1\times10^7$ & 260(20) \\
173 & 16.1\% &$1.7\times10^6$ & 440(25) \\
171 & 14.3\% &$3.4\times10^6$ & 165(10) \\
170 & 3.05\% &$3.0\times10^6$ & 450(30) \\
\end{tabular}
\label{table1}
\end{table}

\begin{table}
\caption{
Listed are the loading rate $R$ and steady state 
population $N_s$ for the various isotopes from Table 1, 
normalized to the values for $^{174}$Yb. For the even 
isotopes, the loading rate scales as the isotopic 
abundance, as expected for a non-enriched thermal source. 
$N_s$ scales roughly as the square root of the abundance. 
The odd isotopes show different behavior due to 
excited-state hyperfine structure, as explained in the text.
}
\begin{tabular}{cccc}
Isotope & Rel. abund. & \multicolumn{1}{c}{$R/R^{174}$} & 
\multicolumn{1}{c}{$N_s/{N^{174}_s}$} \\
\tableline
176 & 0.40 & 0.39(4) & 0.62 \\
174 & 1.00 & 1.00 & 1.00 \\
172 & 0.69 & 0.61(5) & 0.85 \\
173 & 0.51 & 0.055(4) & 0.13 \\
171 & 0.45 & 0.30(2) & 0.26 \\
170 & 0.096 & 0.096(8) & 0.23 \\
\end{tabular}
\label{table2}
\end{table}

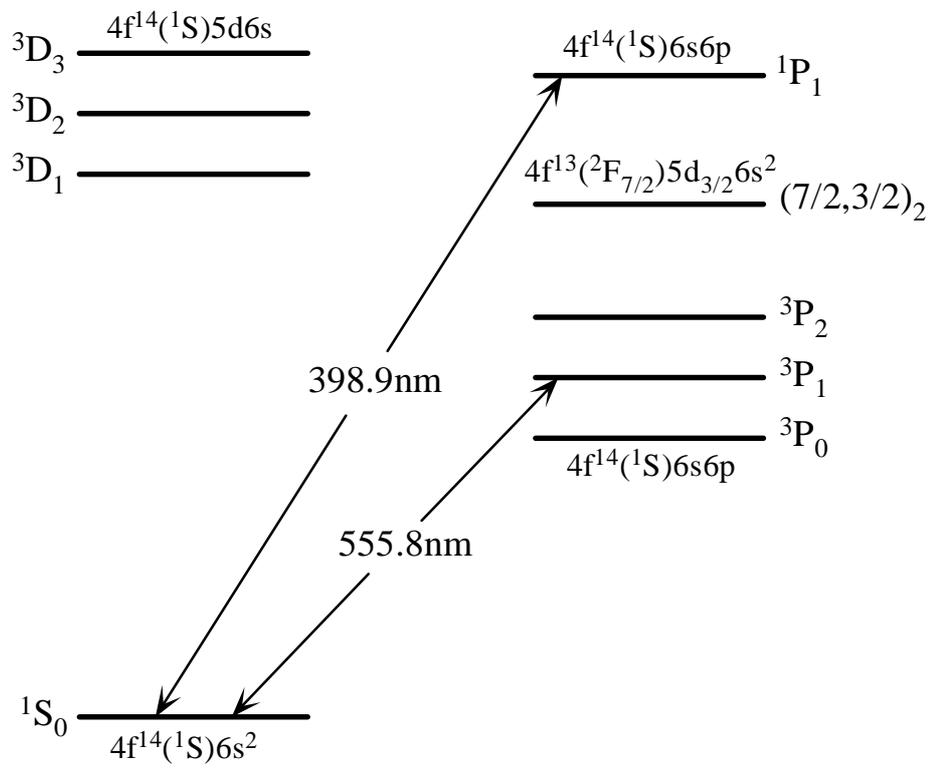
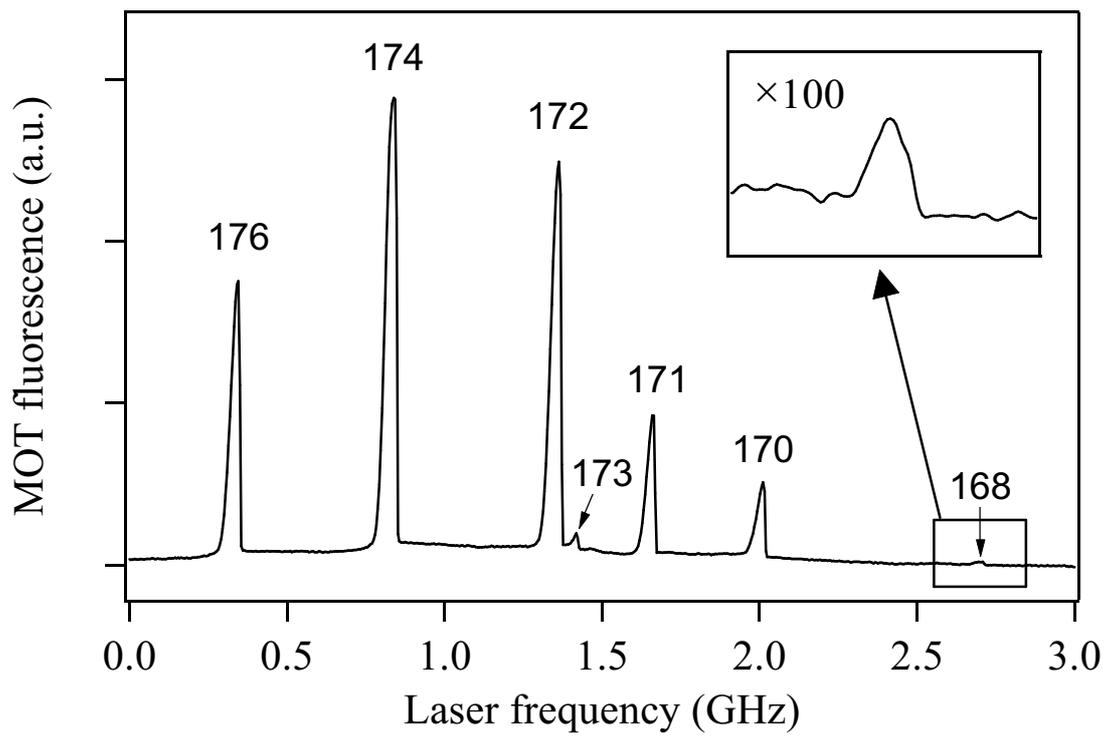
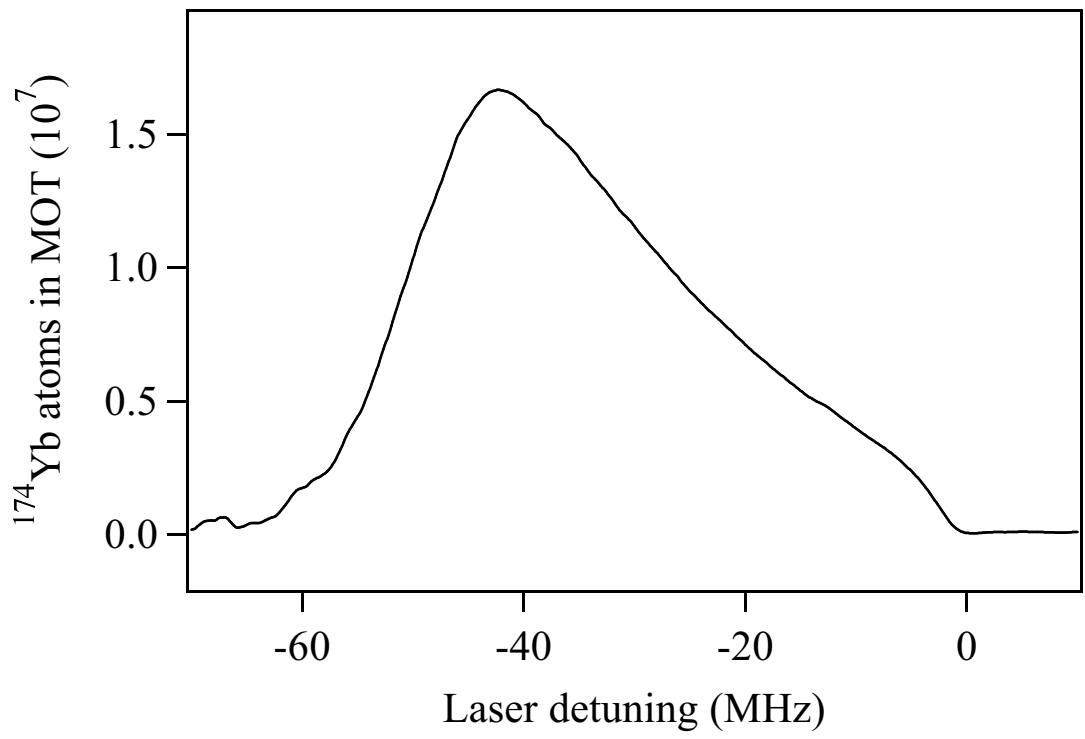
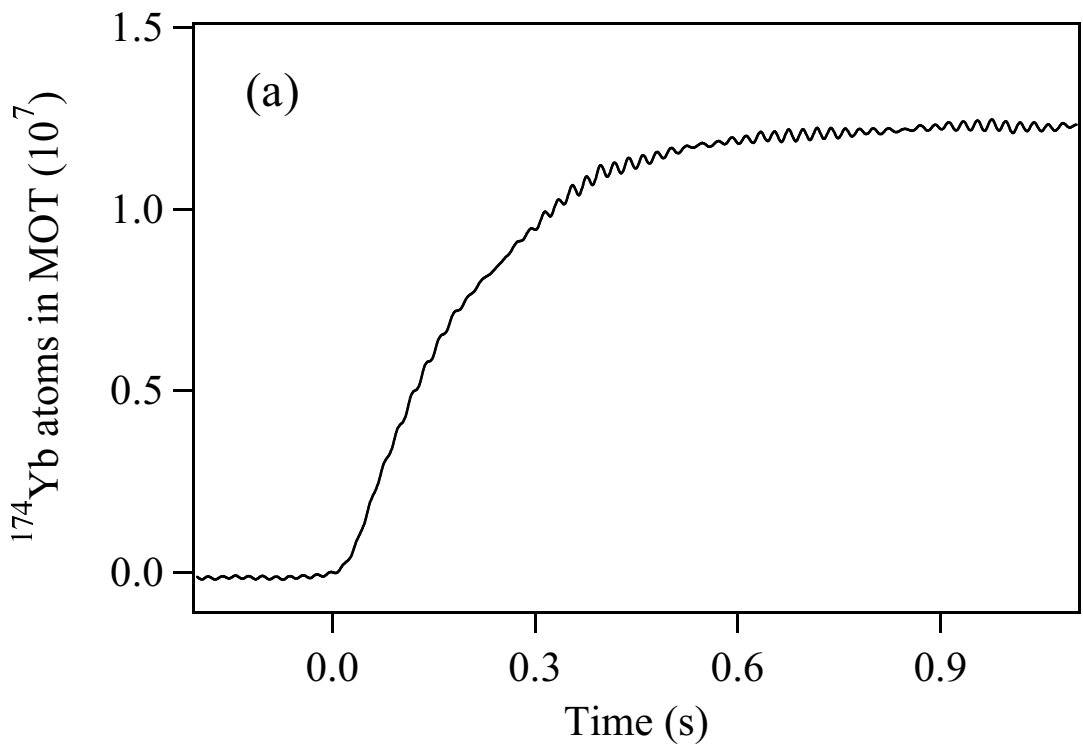
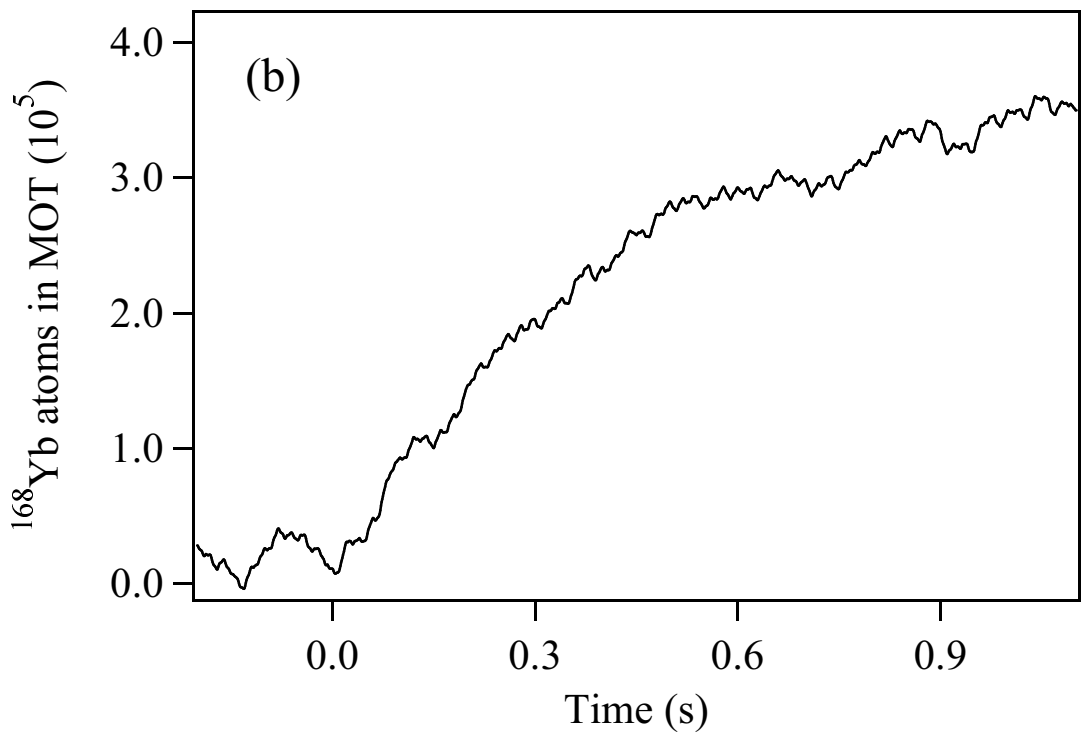
\end{document}